\pdfoutput=1
\documentclass[11pt]{article}
\usepackage[final]{acl}
\usepackage{times}
\usepackage{latexsym}
\usepackage[T1]{fontenc}
\usepackage[utf8]{inputenc}
\usepackage{microtype}
\usepackage{inconsolata}
\usepackage{booktabs}
\usepackage{amsmath}
\usepackage{graphicx}
\usepackage{tikz}
\usetikzlibrary{arrows.meta,backgrounds,positioning,calc,fit,shapes.geometric}

\title{Fenced Citation-Context Retrieval for Case Law: \\ Temporal Leakage and Degree Control Across Two Jurisdictions}

\author{Yao Liu$^{1,2}$ \quad Tien-Ping Tan$^{2}$ \quad Zhilan Liu$^{3}$ \\
  $^1$The Engineering and Technology College, Chengdu University of Technology, Leshan, China \\
  $^2$School of Computer Sciences, Universiti Sains Malaysia, Penang, Malaysia \\
  $^3$Department of Art and Design, The Engineering and Technology College, \\
  Chengdu University of Technology, Leshan, China \\
  Correspondence: \texttt{tienping@usm.my}}

\begin{document}
\maketitle

\begin{abstract}
Prior case retrieval (PCR) aims to identify the precedent cases relevant to the facts of a query case. Incoming citation context, the text with which later cases characterize a case when citing it, is a powerful relevance signal \citep{patil2024pat}, yet it is typically evaluated without a temporal constraint, so the retriever is credited with citations made after the query. We introduce a temporally fenced retriever with no learned parameters that augments BM25 with incoming citation context restricted to citations predating the query, together with a \emph{temporal-admission decomposition} that quantifies the \emph{phantom fraction}: the share of a citation-context gain attributable to citations not known to predate the query. Experiments span two jurisdictions, U.S. federal (CLERC) and European (ECtHR-PCR) case law. On ECtHR-PCR, without any training, the fenced retriever outperforms a strong degree-controlled baseline across the full recall ladder, and a temporal-admission decomposition attributes \textbf{14.9\%} (validation) of an unfenced citation-context gain over BM25 to citations not known to predate the query. Citation-context retrieval must therefore be temporally fenced and degree-controlled before its reported gains can be interpreted.
\end{abstract}

\section{Introduction}
\begin{figure*}[t]
\centering
\includegraphics[width=\textwidth]{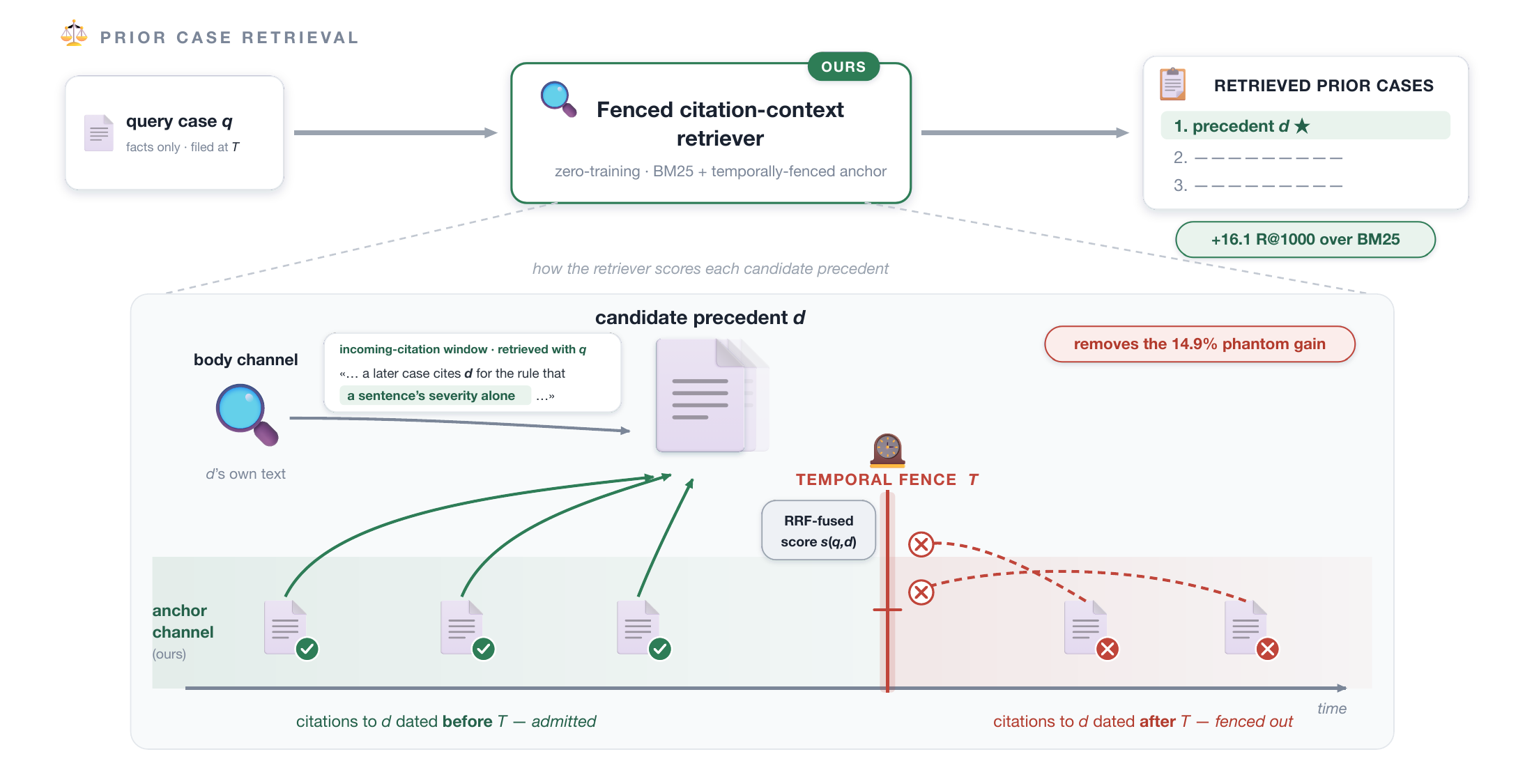}
\caption{Fenced citation-context retrieval. Each candidate precedent $d$ is scored for a query case $q$ by its own text (body channel, BM25) and its \emph{incoming citation context}---the windows with which other cases cite $d$, retrieved with $q$ and RRF-fused---temporally fenced to citers dated before $q$ (green, admitted); later citers (red) are temporal leakage, a $14.9\%$ phantom fraction of the citation-context gain on ECtHR-PCR.}
\label{fig:fence}
\end{figure*}

Prior case retrieval identifies, for the facts of a query case, the earlier decisions on which it relies \citep{joshi2023ucreat,hou2025clerc,santosh2024ecthrpcr}, a core task in precedent-based legal reasoning. Beyond a case's own text, a complementary signal is its \emph{incoming citation context}: the language later cases use to describe a case when they cite it, analogous to the anchor text long exploited in web search \citep{craswell2001anchor}. \citet{patil2024pat} exploit this signal for legal precedent retrieval, aggregating the text surrounding incoming citations to a prior case---its Preceding citation Anchor Text (PAT)---to enrich the case's representation.

This signal, however, is almost always used without a temporal constraint. A candidate's incoming-citation context comes from later cases that cite it, and some of those citing cases are dated after the query; a retriever that uses it without such a constraint is therefore credited in part with citations not known to predate the query---evidence no deployed system could have used. PAT and its successors aggregate citations regardless of date, and IL-PCR states that ``we did not put any temporal constraints on the scraped documents'' \citep{joshi2023ucreat}. A second confound is popularity: when a benchmark's relevance labels are themselves citations, a candidate's raw citation count already predicts relevance, so a citation-context gain may reflect popularity rather than content. Neither the size of this over-credit nor its separation from popularity has been quantified for legal case retrieval.

We address both gaps by building a retriever with no learned parameters that augments BM25 with an incoming-citation signal admitting only citers that predate the query, together with, to measure what such a fence removes, a \emph{temporal-admission decomposition} that splits a naive, unfenced gain into legitimate pre-query evidence and the part from citations not known to predate the query---the \emph{phantom fraction}. On the European Court of Human Rights protocol (ECtHR-PCR), that fraction is \textbf{14.9\%}. Even under the fence, the retriever improves substantially over a tuned BM25 baseline on both jurisdictions and, without any training, outperforms an explicit citation-degree control (for the popularity confound) across the full recall ladder; a published trained system reaches only a comparable R@1000 in an unpaired comparison. The transferable lesson is that a citation-context gain must be temporally fenced and, where relevance labels are citations, degree-controlled before it can be interpreted.

\paragraph{Contributions.}
\begin{enumerate}
\item \textbf{CLERC (US federal, 1.84M docs, BM25 first-stage).} A strictly temporally-fenced anchor channel with no learned parameters. Its primary result is a full-universe policy delta under a byte-invariance-audited mapping: \textbf{+16.1 R@1000 (95\% CI [+14.3, +17.9])} on the 2,850 eligible official test queries (mapped$\rightarrow$fenced, unmapped$\rightarrow$body fallback; strongest reproduced sparse baseline 54.1), with a coverage-conditioned +22.3 on the mapped 2,058. This reproduces, under a mapping that passes a three-input byte-invariance audit (audited-runtime qrel-blindness), the +21.98 previously computed on the superseded label-assisted alignment; it holds at every reported recall cut and on MRR@10, and it clears an explicit fairly-tuned fence-matched citation-degree control (whose dev-optimal degree-fusion weight is zero, so it coincides with BM25; policy LB95 +14.54).
\item \textbf{Temporal-admission estimands, quantified on both protocols.} On ECtHR-PCR (val) we decompose the naive relax-the-fence effect (+4.97 [4.53, 5.43]) into three pre-registered components that sum exactly at the published retrieval depth: a non-prior-date evidence premium (+2.70 [2.36, 3.05]; the clean leakage estimand, similar across both tested caps),\footnote{Date granularity differs by corpus. ECtHR-PCR has day-level dates, so its clean leakage estimand is a \emph{non-prior-date} (same-day-or-later) premium; CLERC has only year-level dates, so its analogous quantity is a \emph{non-prior-year} (same-calendar-year-or-later) admission premium that also admits some legitimately pre-query citers, not a pure future-evidence measurement.} a split-static admission cost (+3.29 [2.57, 4.15]), and an index/retrieval-set effect ($-$1.02 [$-$1.83, $-$0.39]); the two secondary components are depth-sensitive (\S\ref{sec:e1}). We further report a phantom fraction: the +2.70 premium is 14.9\% [12.9\%, 17.1\%] of the full unfenced-over-BM25 gain (equivalently, 54\% of the +4.97 relax-the-fence effect). On CLERC (dev) we turn the admission-premium point estimate into a fence-offset dose--response, where the non-prior-year admission premium rises monotonically (in aggregate point estimates) from 0 at the strict fence to +4.95 with no fence, with cluster-bootstrap simultaneous bands and per-year increments.
\item \textbf{A second-protocol replication with explicit popularity controls (ECtHR-PCR).} Fenced anchor+BM25 reaches macro R@1000 \textbf{79.56}, outperforming BM25 (62.31), degree-only (39.19), window-count-only (42.52), BM25+degree (70.89; primary $\Delta$ +8.68, cluster LB95 +7.62; R@100 $\Delta$ +15.45, LB95 +13.48), and BM25+window-count (71.15), under a single-shot test with pre-registered gates and a provenance audit with zero fence violations. As an unpaired published-rows comparison on the same test protocol, the fenced system, which has no learned parameters, reaches an R@1000 point estimate similar to the strongest published trained system, LeCoPCR-HT \citep[Longformer + concept-guided hybrid training; R@1000 79.39, MAP 13.68;][]{santosh2025lecopcr}, with higher published point estimates at shallow cuts (R@100 51.03 vs 38.62).
\item \textbf{A mechanism observation (exploratory).} CLERC's family/time-matched placebo collapses the gain below body, consistent with candidate-specific target--context pairing mattering there. ECtHR-PCR's placebo permutes score \emph{values} over the roster the query's own term matching had already selected, each candidate keeping its own match count, and retains ${\sim}96\%$ of the gain. \textbf{Scope correction (this version):} that placebo does \emph{not} destroy candidate identity, so its retention bounds score \emph{magnitude} only and separates no structural component; the residual increment (+0.29 to +0.33) is not a candidate-identity effect. The two designs answer different questions, and neither identifies a cross-corpus mechanism.
\end{enumerate}

\section{Related Work}

\paragraph{Anchor / citation-context retrieval.} Anchor text improved early web search \citep{craswell2001anchor}; citation-context has long been used to characterize cited documents in scientific IR \citep{ritchie2008citation}. In law, \citet{patil2024pat} propose PAT, aggregating incoming-citation anchor text to enrich a cited case's representation, and show, with TF-IDF cosine on FIRE IRLeD (2000 prior / 200 queries) and AILA (2914/40), significant MAP gains (JP-PATU 0.5358 vs JP 0.4687). Their abstract's headline, that PAT ``captures certain nuances not captured by the text present in the referenced judgment'', is the beyond-document lesson we build on. During context collection PAT does exclude the query's own judgments and judgments citing the query case, a within-collection contamination control, but not a temporal fence: it places no date constraint between citer and query, so citations not known to predate the query remain admissible. Critically, PAT uses no temporal fence, no placebo control, no degree control, and no query-expansion control; its limitations discuss dataset size and mapping coverage, not deployability or leakage. We add exactly these: strict temporal fencing with leakage quantification, placebo/degree/count controls, and two large-scale protocols. In scientific citation recommendation, \citet{goyal2026profile} build a non-learnable incoming-citation ``profile'' for candidate papers and argue for a strict temporally-inductive evaluation setting, convergent hygiene in a neighboring domain; they do not decompose or quantify the leakage a non-inductive evaluation admits, which is the gap we fill for legal PCR.

\paragraph{CLERC.} \citet{hou2025clerc} release CLERC, whose queries are citing passages with the central citation removed over a corpus of 1.84M federal opinions; they report BM25 R@1000 48.3 and fine-tuned LegalBERT DPR 68.5. We operate in the no-learned-parameter first-stage regime and use these as reference rows, not as our paired comparator. Contemporaneous DALDALL \citep{choi2026daldall} trains dense retrievers on CLERC and COLIEE under a different sampled-query protocol; we treat it as related training work, not a comparator.

\paragraph{ECtHR-PCR.} \citet{santosh2024ecthrpcr} release ECtHR-PCR (15,729 English ECtHR judgments, 1960--2022) expressly to fix realism problems of masked-citation PCR benchmarks: queries are facts sections only, candidate pools contain only cases dated before the query, and splits are chronological. Their baselines are lexical and dense; the paper states its baselines ``focus solely on the text content, neglecting the potential insights from the citation network'', which is the channel we build and fence. LeCoPCR \citep{santosh2025lecopcr} is, to our knowledge, the strongest published system on this test protocol: weak-supervision legal-concept generation (LongT5) augments the query, and a trained Longformer bi-encoder with hybrid noisy-concept training reaches macro R@1000 79.39 (MAP 13.68); its limitations explicitly name the citation network and the temporal dimension as unexploited. We use its published rows as a trained reference comparison: our fenced channel, with no learned parameters, reaches a similar R@1000 point estimate (79.56) and reports higher published point estimates at shallow cuts (R@50 40.16 vs 26.47; R@100 51.03 vs 38.62; MAP 15.82 vs 13.68). \citet{mumford2025citnet} annotate ECtHR citation instances with Article/consideration context and release an analysis tool with context-filtered lookup, with retrieval named as future work; together with the same group's citation-context extraction study \citep{mumford2025goldilocks}, this is concurrent, complementary infrastructure that does not evaluate PCR retrieval or temporal admissibility. (LeCoPCR reports 15,204 cases where the dataset paper reports 15,729; we use the official release and flag the count discrepancy as a preprocessing/reporting difference when reading published rows side by side.)

\paragraph{Other PCR corpora and temporal effects.} IL-PCR \citep{joshi2023ucreat} masks in-pool citation strings corpus-wide and, by design, applies no temporal constraints. COLIEE 2026 suppresses whole citation fragments corpus-wide and lacks reliable document identities: Appendix~\ref{app:coliee} shows suppressed fragments outnumber surviving full-form citations by 314:1 on test and 573:1 on train. Anchor-style enrichment would require importing external identities/graphs, so COLIEE is a negative applicability boundary for same-protocol replication. \citet{premasiri2025llmembed} evaluate zero-shot LLM embedders on four small-pool PCR benchmarks (1--2k candidates) and outperform BM25 there; on CLERC's 1.84M-document pool we observe the opposite direction for zero-shot dense (\S\ref{sec:clerctest}), where pool scale may contribute to the discrepancy alongside model and protocol differences. Temporal degradation of citation-based signals is documented at scale for statute co-citation (UA-StatuteRetrieval; \citealp{ovcharov2026statute}), which also benchmarks co-citation link prediction against a degree/popularity-only control (Adamic--Adar MRR 0.272 vs degree 0.059) under a within-year train/test split: a legal-domain degree control and temporal split, though on statute \emph{link prediction} (recovering a missing citation from a partial set) rather than case-to-case retrieval from query facts. Ovcharov also decomposes an observed temporal citation-retrieval effect through controlled counterfactuals, attributing part of the year-over-year decay to article-composition shift and separately removing evaluation leakage with a temporal train/test split, so the controlled-decomposition move is not ours first; what is distinct here is the target, a retrieval \emph{gain} rather than a temporal decay, split into a clean non-prior-date premium, a legitimate pre-query admission cost, and an index effect with the phantom fraction, for a deployed case-to-case retriever. Beyond legal IR, temporal leakage through date-filtered web retrieval measurably inflates retrospective forecasting evaluations \citep{ellahib2026leakage}, and global-timeline violations change model rankings in recommender evaluation \citep{ji2023leakage}. More broadly, this study belongs to an evaluation-hygiene line in IR that scrutinizes whether reported neural gains survive strong lexical baselines \citep{yang2019neuralhype} and whether zero-training lexical retrieval is systematically underrated \citep{thakur2021beir}. Our contribution is the legal-PCR instantiation: naming the admissible evidence set, fencing it mechanically, and decomposing the retrieval gain an unfenced evaluation over-credits.

\section{Method}

\paragraph{Body channel.} BM25 over case text (CLERC: opinions, dev-tuned $k_1{=}1.2$, $b{=}0.75$; ECtHR-PCR: facts+law, val-tuned $k_1{=}2.0$, $b{=}0.9$ from a pre-registered 25-point grid; both landed on grid edges, and the grids were not extended post hoc).

\paragraph{Anchor channel (no learned parameters).} For each case $t$, we collect windows around every located citation to $t$ in a citing case $c$ (docid \texttt{t|c|i}), retrieve them with the query, aggregate per target, and RRF-fuse with the body channel; the per-query temporal fence (Figure~\ref{fig:fence}) admits only citers dated before the query. Formally, for a query $q$ and candidate case $d$ under a fence threshold $T$,
\begin{align*}
W_T(d) &= \{\, (d,c,i) : \mathrm{date}(c) < T \,\},\\
s_{\mathrm{anc}}(q,d) &= \!\!\sum_{w \in W_T(d)}\!\! \omega(w)\,\mathrm{BM25}(q,w),\\
s(q,d) &= \frac{1}{\kappa + r_{\mathrm{body}}(d)} + \frac{1}{\kappa + r_{\mathrm{anc}}(d)},
\end{align*}
where $(d,c,i)$ is the $i$-th window in which citing case $c$ cites $d$, $T$ is the query date (per-query fence) or the split start (split-static fence), $\omega(w)$ weights each window (unit for ECtHR-PCR; recency-decayed for CLERC), $r_{\mathrm{body}}$ and $r_{\mathrm{anc}}$ are within-channel ranks, and $\kappa$ is the RRF constant ($60$ for CLERC, $30$ for ECtHR-PCR). We say ``no learned parameters'' rather than ``zero-training'': nothing is trained anywhere in the pipeline; retrieval hyperparameters are dev/val-selected from pre-registered grids.

\begin{itemize}
\item \textbf{CLERC variant:} ${\sim}300$-word windows from CAP citation metadata, indexed per citer-year; per-query fence = citer-year $<$ query-source year; same-source skip and citation-family closure; recency decay; RRF $k{=}60$. The deployed admissible graph contains 16.57M anchor edges over 179 per-year Lucene indexes.
\item \textbf{ECtHR-PCR variant:} the dataset itself publishes per-case ISO dates and cited-application lists; windows are $\pm 250$ words (val-frozen from \{100,150,250\}) around digit-bound application-number mentions in the citer's law section (initial 400-pair inline app-number hit rate 84\%; split-static pair-level located coverage 77.2\%). The fence is split-static: anchor artifacts (edges, windows, degree features) use only citers dated before the split start (test $<$ 2018-01-01), so evaluation-split queries, whose citation lists are that split's qrels, contribute nothing to any run. This is enforced \emph{byte-mechanically}: rebuilding the anchor artifacts from a sentineled corpus (test queries' citations and law blanked) must produce hash-identical files, the body inputs must be identical under a citations-only sentinel, and structural asserts run inline. Per-query family exclusion (same application root) applies in every channel. Aggregation = sum of window BM25 scores (val-frozen); RRF $k{=}30$ (val-frozen).
\end{itemize}

\paragraph{Popularity/structure controls (ECtHR-PCR; pre-registered).} Because ECtHR-PCR relevance \emph{is} citation, we require the fenced channel to outperform four popularity/structure controls: a degree-only ranking; BM25+degree fusion (the primary comparator); a window-count-only ranking; and BM25+window-count fusion. We also build a \emph{score-permutation} placebo, pre-registered as \emph{candidate-identity}: among candidates matched on ($\log_2$ fenced degree, $\log_2$ retrieved-window count) it permutes the retrieved windows' BM25 score \emph{values}, each candidate keeping its own window count (a post-hoc, ungated \emph{static-structure} variant instead buckets by static located-window count and samples scores uniformly). It was pre-registered as a pass/fail control; after observing that it retains ${\sim}96\%$ of the gain we treat it as a decomposition instrument instead (this reinterpretation is post-hoc; the measurement and its seed were fixed in advance). \textbf{Scope correction (this version).} It does \emph{not} destroy identity. It permutes over the candidates whose own fenced contexts matched the query, and each keeps its match count, so content-driven match incidence and multiplicity survive in both arms. Its ${\sim}96\%$ retention therefore bounds score \emph{magnitude} given content-selected membership; it separates no structural component, and the residual increment must not be read as a candidate-identity or structure-borne effect. BM25 branches are untouched in all fused controls; all seeds fixed.

\paragraph{Temporal-admission estimands (dev/val only; fence-less variants are never systems).} We pre-registered the following estimands, each a paired within-query contrast under identical scoring; ``clean leakage'' refers only to the ECtHR-PCR non-prior-date premium, the one contrast whose excluded evidence is unambiguously datable relative to the query:

\begin{itemize}
\item \textbf{Non-prior-date premium (ECtHR-PCR; day granularity).} Four arms on val, identical scoring/aggregation/fusion, differing only in citer admission and artifact set: \textbf{a} = split-static fenced (the deployed configuration); \textbf{a$'$} = split-static admission evaluated on the full-artifact index (isolates index construction from admission); \textbf{b} = per-query dynamic admission (citer date $<$ query date); \textbf{c} = unrestricted (any citer date, including evaluation-split queries; family/self exclusions identical in all arms). Then the naive relax-the-fence effect decomposes exactly into a non-prior-date evidence premium $c{-}b$ (the clean leakage estimand), a split-static admission cost $b{-}a'$ (legitimate pre-query citers a split-static fence wrongly excludes), and an index/retrieval-set effect $a'{-}a$:
\begin{equation*}
\underbrace{c-a}_{\text{naive relax}} = \underbrace{(c-b)}_{\text{premium}} + \underbrace{(b-a')}_{\text{admission}} + \underbrace{(a'-a)}_{\text{index}}.
\end{equation*} All arms are evaluated at a fixed retrieval cap (top-$M$), so these are cap-truncated estimands; we report a second depth as labeled sensitivity.
\item \textbf{Phantom fraction (ECtHR-PCR).} The share of the naive unfenced-over-BM25 gain that is non-prior-date evidence,
\begin{equation*}
\phi = \frac{c-b}{\,c-\mathrm{body}\,},
\end{equation*}
with numerator and denominator jointly cluster-resampled in a single stream (fraction computed per resample). Pre-registered denominator-failure rules (report the absolute premium only if the denominator is weak) did not trigger.
\item \textbf{Fence-offset dose--response (CLERC; year granularity).} Admission arms citer-year $< Y(q){+}k$ for $k \in \{0, 1, 2, 5, 10, \infty\}$, with scoring frozen time-blind across all arms; premium($k$) = R@1000($k$) $-$ R@1000($k{=}0$). $k{=}1$ admits same-calendar-year citers (which include pre-query citations), so the curve is a non-prior-year admission dose--response, not a pure future-evidence curve; the clearly-later-year contrast is $k{=}\infty$ $-$ $k{=}1$. We report the cumulative curve, incremental year shells, per-arm admitted-evidence exposure, and, for each of three families (premiums, shells, decay-sensitivity premiums), sup-$t$ simultaneous cluster-bootstrap bands alongside pointwise intervals. A symmetric absolute-decay scoring variant is reported as labeled sensitivity.
\end{itemize}

\section{Experimental Setup}

\paragraph{CLERC.} CLERC is doc-level retrieval over 1.84M US federal opinions; we tune on 2,000 held-out dev queries and evaluate on the official 2,851-query test set \citep{hou2025clerc}. The fenced channel needs each query's source-case identity and date. Our original run took these from a label-assisted alignment, so we re-derived them end-to-end with a \textbf{post-exposure, pre-specified mapping that passes a three-input byte-invariance audit}: real, permuted, and fake qrels each produce byte-identical mapping artifacts, establishing audited-runtime qrel-blindness (invariance to those three inputs, not a proof that no cached label-derived channel exists).\footnote{Calibrated on the official generation-train split at a frozen operating point (matching threshold 9): precision 100\% (3,853/3,853), date availability 87.29\% (4,059/4,650), conditional recall 94.92\% (3,853/4,059), unconditional recall 82.86\% (3,853/4,650). The map agrees with the superseded label-assisted alignment on 100.0\% of source years and 99.95\% of source IDs over their 2,027-query intersection, evidence that it is not label-shaped.} Coverage is disclosed: raw mapping covers 2,059/2,851, and one benchmark self-reference (its gold document verbatim-contains the query) is excluded by a frozen allowlist, leaving an \textbf{eligible universe of 2,850} with \textbf{2,058 mapped (72.21\%)}; we never quote 2,058/2,851 as coverage. Our \textbf{primary estimand, frozen before the audit's single test-qrel read, is the policy $\Delta$R@1000} (mapped$\rightarrow$fenced, unmapped$\rightarrow$body fallback) against body BM25(1.2,0.75) on the 2,850, with a source-cluster paired bootstrap (mapped clusters by source, unmapped as singletons; $B{=}10{,}000$; ratio-of-sums; two-sided 95\% CI, 2.5th percentile as decision statistic). All configurations were dev-frozen and the audit changed no parameters; its only new degrees of freedom, the map and the frozen policy/estimand/decision rule, were fixed before its qrel read.\footnote{The degree control, dense comparator, and provenance audit are separately frozen analyses (chronology in \S\ref{sec:limitations}). Dose--response: dev, 2,000 source clusters, $B{=}10{,}000$, seed frozen in advance; replayed published endpoints exactly (body 84.35 / fenced 88.35 / unfenced 93.30 / decay arms 93.45/96.40).}

\paragraph{ECtHR-PCR.} Official protocol \citep{santosh2024ecthrpcr}: 15,729 cases, with a per-query candidate pool of all cases strictly before the query date and facts-only queries; the \textbf{test set is 3,231 gold-bearing queries (2018--2022; 38,759 gold pairs)}; metrics are \textbf{macro-averaged} per the dataset paper, and we report R@\{50,100,500,1000\}+MAP. Tuning is two-stage on val (2015--2017; 2,186 queries): freeze BM25 (25-point grid), then freeze the anchor configuration ($\leq$36-point grid). \textbf{Reproduction gate:} our tuned BM25 reaches 62.31 vs the published 60.38 (+1.93; a \emph{stronger} baseline is conservative for our deltas). \textbf{Significance:} exact-duplicate facts are common (91 clusters covering 500/3,231 test queries), so the primary gate uses a cluster paired bootstrap over union(application-root, exact-normalized-facts) clusters (2,822 test units; $B{=}10{,}000$). Test was run once, after all pre-registered byte-equality and structural gates passed. The temporal-admission decomposition: val, clusters recomputed on val (2,036 units over 2,186 queries), $B{=}10{,}000$; the frozen a and c arms replay the published 82.47/87.44/+4.97 exactly (rebuilt and reproduced on a second machine from sha-pinned inputs).

\paragraph{Controls \& audits.} CLERC: family/time-matched placebo; doc2query control (unsupported: weak opponent, disclosed); index-fidelity check (official BM25 reproduced exactly, 0.4826); adversarial provenance audit. ECtHR-PCR: the degree/count controls and score-permutation placebo above; flip-level provenance audit re-verifying every contributing window against raw artifacts; same-facts-citer sensitivity; window-precision human audit (100-sample, seed-fixed).

\section{Results}

We report CLERC first (two dev diagnostics followed by the audited test result), then the single-shot ECtHR-PCR confirmation and its temporal-admission decomposition, closing with what transfers across the two protocols. Throughout, dev/val numbers are diagnostics; the two test reads carry the headline claims.

\subsection{CLERC dev: the gain survives a family/time-matched placebo}
\label{sec:clercdev}
The fenced anchor channel lifts dev R@1000 by \textbf{+9.10 points} (body 0.8435 $\rightarrow$ 0.9345), and a family/time-matched placebo is consistent with that lift being real content rather than an artifact of the contexts alone: permuting the target each real context is attached to sends the placebo \emph{below} body (0.8245), while the matched real-context arm holds at 0.9315, a miss@1000 relative reduction of \textbf{0.6097} (one-sided LB95 0.5658; source-cluster sign +214/$-$0). A doc2query control does not qualify as an opponent (best 0.8550 $\approx$ body 0.8525), so we report it as unsupported rather than validating.

\subsection{CLERC dev: removing the fence over-credits, and the over-credit grows as the year fence is relaxed}
\label{sec:dose}
\begin{figure}[t]
\centering
\includegraphics[width=\linewidth]{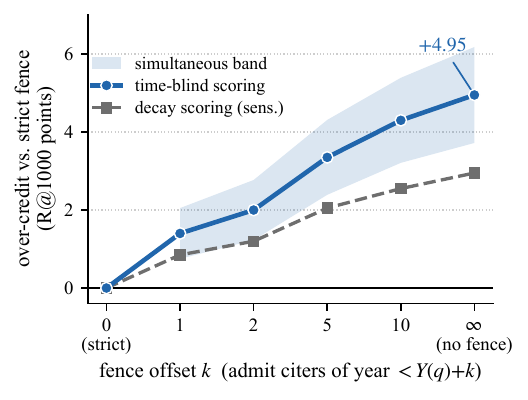}
\caption{CLERC dev fence-offset dose--response: relaxing the year fence over-credits the anchor channel, monotonically in aggregate point estimates, from 0 to +4.95 R@1000 (time-blind scoring; shaded band = sup-$t$ simultaneous cluster bootstrap, $B{=}10{,}000$; dashed = recency-decay-scoring sensitivity, ending at +2.95). $k{=}1$ admits same-calendar-year citers (some pre-query); the clearly-later contrast $k{=}\infty-k{=}1$ is +3.55. Per-arm values and bands in Appendix~\ref{app:dose}.}
\label{fig:dose}
\end{figure}

A fence-less CLERC variant would over-report a non-prior-year (same-calendar-year-or-later) admission premium of +2.95 R@1000 points under recency-symmetric decay scoring, rising to +4.95 under time-blind scoring, each measured against its own matched fenced configuration (we do not impute these to PAT). Turning that endpoint into a dose--response (Figure~\ref{fig:dose}), by admitting citers one year band at a time with scoring frozen time-blind across arms, the premium climbs monotonically (in aggregate point estimates) from 0 to +4.95, and every relaxed arm's simultaneous band excludes zero. The decay-scoring curve is likewise monotone, ending at +2.95.

We read the curve's \emph{shape} cautiously. The five incremental year shells sum exactly to the endpoint, but they are ranking contrasts under RRF, not additive causal contributions of independent time bands. The same-calendar-year shell (+1.40) is the largest, yet it mixes citers that precede and follow the query within one calendar year and leads the Y+2..Y+4 shell by only 0.05, so we claim no ``near-future dominates'' effect; the Y+10-and-later shell is positive pointwise ([0.10, 1.25]) but its five-shell simultaneous band ([$-$0.08, 1.38]) does not exclude zero. The clearly-later-year contrast ($k{=}\infty$ $-$ $k{=}1$) is +3.55 points. Admitted evidence also grows with each arm (median admitted hits 6,250 at $k{=}0$ $\rightarrow$ 7,551 at $k{=}\infty$), most sharply at the final arm, where the premium increment of +0.65 is among the smallest; so temporal reach and evidence mass move together, and the curve's shape alone licenses no mechanism story.

\subsection{CLERC test: an audited post-exposure replication}
\label{sec:clerctest}
On the official CLERC test set the fenced policy substantially lifts recall: \textbf{policy R@1000 = 70.21 vs body BM25's 54.11, $\Delta$ = +16.11 points (95\% CI [+14.28, +17.95])} (cluster bootstrap per \S4; one-sided LB95 +14.54; miss@1000 relative recovery RR = 0.351 [0.316, 0.385]; query-level gains/losses 499/40; cluster sign +167/$-$6/=120 among mapped clusters). We label the result carefully. The CLERC test set was exposed during the original label-assisted run, so this is a transparent post-exposure estimate re-derived under a provenance audit, not a fresh blind confirmation; ECtHR-PCR (\S\ref{sec:echrtest}) is the single-shot clean confirmatory experiment, and every CLERC test-qrel read is enumerated in Table~\ref{tab:reads}. The number is the policy estimand on the eligible universe of 2,850 official test queries (2,851 minus one benchmark self-reference; \S4): the 2,058 mapped queries are served by the fenced channel, the 792 unmapped fall back to the body run, and the whole policy is compared against body BM25(1.2,0.75) on the same 2,850. It is deployable and coverage-honest: the 792 fallback queries contribute structural zeros to the numerator while remaining in the denominator.

A secondary, coverage-conditioned diagnostic (mapped queries only; 2,058/2,850, 72.21\%; not the eligible-universe policy estimand and not the paper's headline result) gives fenced 76.82 vs body 54.52, $\Delta$ = +22.30. This reproduces, under the mapping that passed the three-input byte-invariance audit, the +21.98 previously computed on the superseded label-assisted alignment (2,493 queries; \S\ref{sec:limitations} chronology). Body recall on the mapped subset (54.52) is nearly identical to the full eligible universe (54.11), so the mapped subset is not an easier slice, and the exact identity $\Delta_{\text{policy}} = (2{,}058/2{,}850) \times \Delta_{\text{mapped}}$ links the two rows. The gain holds at every operating point on both universes (Table~\ref{tab:ladder}); $\Delta$R@100 (+0.219) $\approx$ $\Delta$R@1000 (+0.223) on the mapped arm, so essentially the full retrieval-recall gain is available within a top-100 reranking pool. The adversarial provenance audit of the original run's 593 body-miss$\rightarrow$fenced-hit flips (computed on the superseded alignment, whose per-query fenced payloads the audited rerun replays exactly for all agreeing queries) records 11,193 gold-context contributions, 173 family-flagged contexts, and low query$\leftrightarrow$context Jaccard overlap (mean 0.0021, p90 0.0087); removing all family-flagged contexts collapses only \textbf{9/593} flips (1.5\% rounded), with verdict GENUINE.

Against a controlled zero-shot dense comparator (single-shot bge-large-en-v1.5, MaxP over the official CLERC passage collection), the fenced system outperforms zero-shot dense and its BM25 hybrid: dense 0.5227 and frozen BM25+dense RRF($k{=}60$) hybrid 0.5965 vs fenced \textbf{0.7726} (paired fenced$-$dense +24.99 R@1000 [22.76, 27.28]; fenced$-$hybrid +17.61 [15.79, 19.48]). These paired dense contrasts were computed on the superseded label-assisted alignment (2,493 queries) and are labeled as such; given the audit's 99.95\% source-ID agreement and the near-identical subset recall (Table~\ref{tab:ladder}), we quote them at that scope rather than re-slicing. Zero-shot dense underperforms BM25 (dense$-$sparse $-$3.01 [$-$4.93, $-$1.07]), consistent with CLERC's own BGE-vs-BM25 direction (42.4 vs 48.3); we make no trained-dense comparison (Table~\ref{tab:consistency}).

The CLERC gain also survives an explicit, fairly-tuned degree control. We compute per-query temporally-fenced citation degree from the same admissible anchor graph the deployed system retrieves from (closure-filtered, near-duplicate-dropped, rule-filtered; all five build counters replayed exactly against the frozen build manifest), and, closing the tuning-asymmetry limitation of the earlier system-weighted control, we give the control its own dev-tuned configuration (a frozen 51-config grid over distinct-citer degree signals: decayed/undecayed, fusion weight $\beta$ incl.\ body-only and degree-only endpoints, RRF $k$; winner selected on dev, then one test evaluation under the frozen policy estimand). No nonzero configuration in the frozen grid out-scored BM25 body on dev: the dev-optimal fusion weight on the degree signal is zero ($\beta{=}0$ leads the best nonzero degree configuration by a single dev query, 1,687 vs 1,686/2,000, and the system-nominal $\beta{=}0.25$ by 19, sitting 0.95 points below body, 0.8340 vs 0.8435). We therefore do not claim the degree signal is information-free, only that no in-grid fusion of it out-scored BM25 on dev; the dev-selected fairly-tuned degree control coincides with BM25 body, so the fenced channel clears it by the same margin as over BM25: \textbf{policy +16.11 R@1000 [+14.28, +17.95]} (one-sided LB95 +14.54; coverage-conditioned +22.30 on the 2,058 mapped queries [+20.06, +24.62]; mapped cluster sign +167/$-$6/=120). The pre-specified reading is that fenced retrieval has higher recall than the fairly-tuned degree control, which does not fully account for the observed difference. This closes the \emph{procedural} tuning asymmetry the earlier system-weighted control left open, but it yields a numerically \emph{weaker} comparator than the nonzero system-weighted control; the popularity rebuttal therefore does not rest on this control's margin. It rests instead on the family/time-matched placebo (which drops the CLERC gain \emph{below} body; \S\ref{sec:clercdev}) and on degree-only scoring far below body (9.35 vs 55.27). The grid, selection rule, seed, and single test-qrel read were sealed before the read (local no-clobber sentinel), and dev$\leftrightarrow$test source overlap is zero. The earlier system-weighted control (fenced 77.26 vs 57.36 on the superseded 2,493-query alignment, +19.90, LB95 +18.20) is retained as superseded context; its ingredients (that control +2.09 over body, LB95 +1.26; degree-only 9.35 and count-only 9.55 vs body 55.27) already showed CLERC's fenced popularity prior is weak, unlike ECtHR-PCR's degree-only 39.19. A separately-labeled popularity-family (window-count) sensitivity, tuned by the identical procedure, likewise does not out-score the degree family on dev.

\begin{table*}[t]
\centering
\small
\begin{tabular}{lrrr@{\hskip 12pt}rrr}
\toprule
 & \multicolumn{3}{c}{Policy universe (2,850)} & \multicolumn{3}{c}{Mapped arm (2,058, coverage-cond.)} \\
\cmidrule(lr){2-4}\cmidrule(lr){5-7}
Metric & Body elig. & Policy (ours) & $\Delta$ & Body mapped & Fenced mapped & $\Delta$ mapped \\
\midrule
R@1    & 0.0014 & \textbf{0.0404} & +0.039 & 0.0015 & \textbf{0.0554} & +0.054 \\
R@5    & 0.0961 & \textbf{0.1596} & +0.064 & 0.1050 & \textbf{0.1929} & +0.088 \\
R@10   & 0.1533 & \textbf{0.2368} & +0.084 & 0.1623 & \textbf{0.2779} & +0.116 \\
R@20   & 0.1944 & \textbf{0.3214} & +0.127 & 0.2080 & \textbf{0.3839} & +0.176 \\
R@100  & 0.3319 & \textbf{0.4902} & +0.158 & 0.3455 & \textbf{0.5646} & +0.219 \\
R@1000 & 0.5411 & \textbf{0.7021} & +0.161 & 0.5452 & \textbf{0.7682} & +0.223 \\
MRR@10 & 0.0422 & \textbf{0.0933} & +0.051 & 0.0450 & \textbf{0.1158} & +0.071 \\
\bottomrule
\end{tabular}
\caption{Official CLERC test metric ladder under the audited mapping (descriptive ladder of the same audited runs; the pre-specified statistic is the R@1000 policy $\Delta$). Left: policy universe (2,850; body vs mapped$\rightarrow$fenced/unmapped$\rightarrow$body policy). Right: mapped arm (2,058, coverage-conditioned). BM25 official (0.9,0.4) on the full 2,851 test = R@1000 0.48264, reproducing the published 0.483, the audited rerun's machinery gate.}
\label{tab:ladder}
\end{table*}

\subsection{ECtHR-PCR test: at zero training, the fenced channel outperforms every popularity control across the recall ladder}
\label{sec:echrtest}
On the single-shot ECtHR-PCR test (macro, 3,231 queries; Table~\ref{tab:echr}) the fenced channel clears every pre-registered popularity control and, at zero training, reaches a published R@1000 point estimate similar to the strongest trained system's. Because relevance here \emph{is} citation, the pre-registered primary comparator is not plain BM25 but BM25+degree, itself strong (70.89, versus degree-only 39.19); the fenced channel still exceeds it by \textbf{+8.68 (cluster LB95 +7.62)}, and by more at shallower cuts (R@100 $\Delta$ +15.45, LB95 +13.48). The strongest published trained system we are aware of, LeCoPCR-HT (a trained Longformer with concept-guided query augmentation and hybrid training), reports R@1000 79.39 and MAP 13.68; our system, with no learned parameters, reaches a similar R@1000 point estimate (79.56) and is higher at every shallower cut and on MAP (R@50 40.16 vs 26.47; R@100 51.03 vs 38.62; MAP 15.82 vs 13.68). This is an unpaired published-rows comparison; we did not re-run their system. Note also that our tuned BM25 (62.31) already exceeds the published BM25 on this protocol (60.38) by 1.93 R@1000---more than ten times the 0.17-point system gap---so the R@1000 match reflects a stronger first stage as much as the anchor channel and should be read as context, not a parity test. The flip-level provenance audit re-verifies all 20,921 windows contributing to the 3,484 golds the fenced system recovers over BM25+degree: zero temporal, self, query-family, or target-family violations, and dropping windows whose citer shares the query's exact facts changes R@1000 by only 0.04. A manual precision grading of the seed-frozen 100-window sample finds 98/100 true citation contexts; both failures are 1--2-digit application serials whose located match is a date or Convention article number.

\begin{table}[t]
\centering
\scriptsize
\setlength{\tabcolsep}{2.4pt}
\begin{tabular}{lrrrrr}
\toprule
Run / control & R@50 & R@100 & R@500 & R@1000 & MAP \\
\midrule
\textbf{Fenced+BM25 (ours)} & \textbf{40.16} & \textbf{51.03} & \textbf{71.68} & \textbf{79.56} & \textbf{15.82} \\
Candidate-identity placebo & 39.32 & 50.09 & 71.06 & 79.27 & 15.43 \\
BM25 + window-count & 28.56 & 36.19 & 61.05 & 71.15 & 9.05 \\
BM25 + degree (primary) & 28.30 & 35.58 & 59.91 & 70.89 & 8.97 \\
BM25 (tuned) & 23.38 & 29.47 & 48.51 & 62.31 & 10.07 \\
Window-count only & 11.67 & 15.01 & 31.36 & 42.52 & 4.62 \\
Degree only & 11.62 & 15.07 & 30.37 & 39.19 & 4.56 \\
\midrule
\emph{LeCoPCR-HT}$^{\dagger}$ & \emph{26.47} & \emph{38.62} & \emph{67.14} & \emph{79.39} & \emph{13.68} \\
\emph{BM25 / dense}$^{\dagger}$ & --- & --- & --- & \emph{60.38 / 67.31} & --- \\
\bottomrule
\end{tabular}
\caption{ECtHR-PCR official test (macro \%, single shot; systems and pre-registered controls). Primary gate: the fenced system vs BM25+degree. The score-permutation placebo (pre-registered as candidate-identity) is a decomposition instrument that bounds score magnitude only; it does not destroy identity (\S3). $^{\dagger}$Published reference rows \citep{santosh2025lecopcr,santosh2024ecthrpcr}; LeCoPCR-HT is trained (Longformer + concept-guided hybrid training); unpaired comparison.}
\label{tab:echr}
\end{table}

\subsection{ECtHR-PCR val: decomposing what a fence-less evaluation over-credits}
\label{sec:e1}
\begin{figure}[t]
\centering
\includegraphics[width=0.94\linewidth]{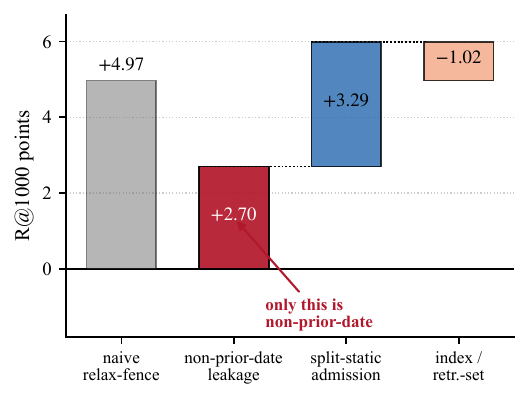}
\caption{Decomposing the naive relax-the-fence effect on ECtHR-PCR val (top-$M{=}5{,}000$). The naive +4.97 R@1000 splits exactly into a non-prior-date evidence premium (+2.70; the only potentially leaking component), a split-static admission cost (+3.29; legitimate pre-query citers a split-level fence must exclude), and an index/retrieval-set effect ($-$1.02). At this depth the largest single component of the naive over-report is legitimate pre-query admission, not leakage, though this ordering is depth-dependent: at top-$M{=}20{,}000$ the non-prior-date premium is instead the largest (Table~\ref{tab:e1}). Exact CIs and a second depth in Table~\ref{tab:e1}.}
\label{fig:decomp}
\end{figure}

Removing the fence inflates the ECtHR-PCR score, and the largest single piece of that inflation is legitimate pre-query admission, not leakage; we can say exactly how the pieces add up. The fence-less variant (any-date citers, including evaluation-split queries) scores macro R@1000 = 87.44 against the fenced 82.47, a naive relax-the-fence effect of \textbf{+4.97 [4.53, 5.43]} (paired cluster bootstrap). Figure~\ref{fig:decomp} and Table~\ref{tab:e1} split this composite into the four pre-registered arms (\S3), whose contrasts sum to it by construction: $(c{-}b) + (b{-}a') + (a'{-}a) = c{-}a$, i.e.\ $2.70 + 3.29 - 1.02 = 4.97$.

\begin{table*}[t]
\centering
\small
\begin{tabular}{lrr}
\toprule
Arm / contrast & top-$M{=}5{,}000$ & top-$M{=}20{,}000$ \\
\midrule
body (BM25) & 69.38 & 69.38 \\
a = split-static fenced (deployed) & 82.47 & 81.69 \\
a$'$ = split-static, full idx & 81.45 & 82.18 \\
b = per-query dynamic & 84.75 & 84.38 \\
c = unrestricted & 87.44 & 86.87 \\
\midrule
\textbf{c$-$b: non-prior-date premium (PRIMARY)} & \textbf{+2.70 [2.36, 3.05]} & \textbf{+2.49 [2.15, 2.85]} \\
b$-$a$'$: split-static admission cost & +3.29 [2.57, 4.15] & +2.21 [1.88, 2.57] \\
a$'$$-$a: index/retrieval-set effect & $-$1.02 [$-$1.83, $-$0.39] & +0.49 [0.26, 0.72] \\
\bottomrule
\end{tabular}
\caption{Temporal-admission decomposition (ECtHR-PCR val, macro R@1000; paired cluster bootstrap, 2,036 clusters, $B{=}10{,}000$; two-sided 95\% CIs). Primary depth top-$M{=}5{,}000$ (published); top-$M{=}20{,}000$ as labeled sensitivity.}
\label{tab:e1}
\end{table*}

Three readings, in decreasing strength. First (primary, and similar across both tested caps): genuinely non-prior-date evidence, meaning citation windows dated on or after the query date, inflates the unfenced evaluation by \textbf{+2.70} points, and the estimate is similar across the two tested caps (+2.49 at 20k; both cap-saturated). This is the clean leakage number, and it is substantially smaller than the naive +4.97: the largest single component of the naive relax-the-fence effect is not leakage but the admission cost of the split-static fence, the legitimate pre-query citers that a byte-mechanically verifiable split-level fence must exclude (+3.29 at the published depth, partially offset by the $-$1.02 index/retrieval-set effect). Second: the split between admission cost and the index/retrieval-set effect is depth-sensitive (the index effect flips sign at 20k), so we license no depth-invariant story for the components; only their sum with $c{-}b$ is similar across the two tested caps. Third (disclosure): the per-query dynamic arm b exceeds the deployed split-static arm a by +2.28 [1.84, 2.72], a composite contrast (admission and artifact set differ) reported for completeness, not as a deployable-signal cost estimate. All arms saturate the retrieval cap on every query, so all rows are cap-truncated estimands (\S\ref{sec:limitations}).

\paragraph{Phantom fraction.} Jointly resampling numerator and denominator, $(c{-}b)/(c{-}\text{body})$ = \textbf{14.9\% [12.9\%, 17.1\%]}: about one seventh of the \emph{naive} unfenced-anchor gain over BM25 on val is evidence dated on or after the query date. Both pre-registered denominator-strength rules pass (denominator 18.06 points, LB95 16.9; zero non-positive replicates).

\subsection{The lesson transfers, but the mechanism differs by protocol}

Across CLERC and ECtHR-PCR, unfenced incoming-citation enrichment over-credits itself by a measurable non-prior-date (ECtHR-PCR) or non-prior-year (CLERC) admission premium. Under strict temporal fencing, the gain remains large in both protocols and clears an explicit citation-degree control in both: CLERC's policy estimand gains +16.11 R@1000 over tuned BM25 on the eligible 2,850 (coverage-conditioned +22.30 on mapped queries) and clears the fairly-tuned degree control (dev-optimal degree-fusion weight zero $\rightarrow$ the control coincides with BM25; policy LB95 +14.54); ECtHR-PCR gains +8.68 macro R@1000 over BM25+degree (cluster LB95 +7.62). The deployability lesson therefore transfers: \textbf{citation-context retrieval must be temporally fenced and degree-controlled before its gains are interpretable.}

The residual mechanism differs by protocol. In CLERC, placebo controls collapse the gain, and against the explicit degree control the fenced residual remains large (policy +16.11 / mapped +22.30 over a fairly-tuned degree control whose dev-optimal degree weight is zero; under the superseded system-weighted control, +19.90, with BM25+degree improving over body by only +2.09), consistent with a candidate-specific-content explanation. In ECtHR-PCR, the pre-registered score-permutation placebo retains about 96\% of the gain (a post-hoc static-structure variant agrees), leaving only a small residual increment (+0.29 to +0.33 R@1000; cluster-bootstrap 95\% CIs [+0.19, +0.39] pre-registered and [+0.24, +0.43] static-structure, both excluding zero). \textbf{Scope correction (this version):} because that permutation runs over the roster the query's own term matching had already selected and preserves each candidate's match count, the retention bounds score \emph{magnitude} given content-selected membership; it does \emph{not} license reading the gain as structure-borne rather than candidate-specific. The two placebo designs are not matched across corpora, so we read the contrast as observational. This is consistent with ECtHR-PCR's smaller, denser citation setting (median 14 fenced third-party citers per gold vs sparse citers over 1.84M US opinions), but we treat that as an explanatory hypothesis rather than a proved general rule.

\begin{table*}[t]
\centering
\small
\begin{tabular}{p{0.21\textwidth}p{0.37\textwidth}p{0.34\textwidth}}
\toprule
 & CLERC (US federal, 1.84M docs) & ECtHR-PCR (ECtHR, 15.7k cases) \\
\midrule
Fenced system (no learned params) & anchor+BM25 RRF, per-query year fence & split-static fenced anchor+BM25 RRF \\
Fenced gain & policy +16.11 R@1000 vs tuned sparse (eligible 2,850; CI [14.28, 17.95]; RR 0.351); coverage-conditioned +22.30 (mapped 2,058) & +8.68 macro R@1000 vs BM25+degree (LB95 7.62) \\
Admission premium (dev/val only) & non-prior-year: dose--response 0 $\rightarrow$ +4.95 (time-blind), monotone in aggregate; decay-scoring endpoint +2.95 & non-prior-date: +2.70 [2.36, 3.05] clean premium within naive +4.97; phantom fraction 14.9\% \\
Popularity/degree control & explicit, \textbf{fairly dev-tuned} (fence-matched): dev-optimal degree weight 0 $\rightarrow$ control = BM25; fenced $-$ control = policy +16.11 / mapped +22.30 (LB95 +14.54); superseded system-weighted control 57.36, fenced $-$ control +19.90 & explicit: degree-only 39.19 / BM25+degree 70.89 / window-count-only 42.52 / BM25+window-count 71.15 \\
Placebo outcome & collapses (consistent with candidate-specific pairing) & retains ${\sim}96\%$ (bounds score magnitude given the content-selected roster; separates no structural component; residual +0.29--0.33) \\
Provenance audit & 9/593 flip collapses after family-flag removal (1.5\% rounded); gain after family-flag removal +21.62 & 0 violations / 20,921 windows; window precision 98/100; same-facts $\Delta$0.04 \\
Trained reference (published rows) & LegalBERT-DPR 68.5 (no paired comparison; reproduction attempt disclosed in \S\ref{sec:limitations}) & dense bi-encoder 67.31; \textbf{LeCoPCR-HT 79.39 vs ours 79.56} (no learned parameters; shallow cuts favor ours) \\
Zero-shot dense comparator & BGE-large-en-v1.5 MaxP: R@1000 52.27 aligned; fenced $-$ dense +24.99 [22.76, 27.28]; hybrid 59.65, fenced $-$ hybrid +17.61 [15.79, 19.48] & not run as controlled head-to-head; published dense rows remain reference only \\
\bottomrule
\end{tabular}
\caption{Two-protocol consistency. CLERC dense rows are single-shot zero-shot BAAI/bge-large-en-v1.5 comparators over the official CLERC passage collection; paired claims use the aligned 2,493-query subset with source-cluster bootstrap (314 clusters, $B{=}10{,}000$), while full 2,851-query dense/hybrid values are context only. Published trained rows (CLERC LegalBERT-DPR 68.5; ECtHR-PCR dense 67.31 and LeCoPCR-HT 79.39) are reference rows; we make no paired trained comparison.}
\label{tab:consistency}
\end{table*}

\section{Limitations}
\label{sec:limitations}

\begin{itemize}
\item \textbf{Borrowed mechanism}: the anchor-text idea is PAT's; our contribution is the fenced deployability measurement, the temporal-admission estimands, the controls, and the cross-protocol replication.
\item \textbf{CLERC coverage and the mapping audit (chronology)}: our original run obtained query source/date metadata from a label-assisted alignment that read test qrels during construction, a provenance defect we later repaired end-to-end with a post-exposure pre-specified test-qrel-free audit: a mapper passing a three-input byte-invariance audit (real/permuted/fake qrels each yield byte-identical artifacts; audited-runtime qrel-blindness, not absolute freedom from every pre-baked channel), calibrated on official train-split labels; it agrees with the superseded alignment on 100.0\% of source years and 99.95\% of source IDs (2,027-query intersection). Two post-failure specification changes were adjudicated and frozen during the audit and are disclosed as such: the calibration objective's feasible-recall denominator, and the exclusion of one benchmark self-reference query (its gold verbatim-contains the query). The paper's primary is the full-eligible-universe policy estimand (2,850), so no aligned-subset number is extrapolated; the mapped/unmapped difficulty gap is small (body R@1000 54.52 vs 53.03).
\item \textbf{Chronology of CLERC test-qrel reads (Table~\ref{tab:reads})}: we enumerate every read rather than assert cleanliness in prose. No retrieval configuration was changed after seeing any test result; each analysis's grid/estimand/seed was frozen before its own read. Preserving a \emph{fully confirmatory} CLERC status would require a fresh blind evaluation on held-out queries; we do not claim it, and foreground ECtHR-PCR (\S\ref{sec:echrtest}) as the single-shot clean confirmation.
\item \textbf{Cap-truncated estimands}: every decomposition arm saturates its retrieval cap on every query, at both depths, so the decomposition is an estimand of the capped evaluation (as published), not a depth-converged quantity; the primary contrast $c{-}b$ is similar across both tested caps, the two secondary components are not.
\item \textbf{Dose--response interpretation}: the dose--response arms vary temporal reach and admitted evidence mass together; the monotone curve quantifies over-crediting as a function of fence relaxation but does not identify \emph{why} each shell contributes (an exposure-matched control is future work). Aggregate monotonicity is a point-estimate property; per-query rank flips occur in both directions.
\item \textbf{ECtHR-PCR supply cost of the split-static fence}: pair-level fenced coverage drops from 94.9\% (per-query dynamic fence) to 80.0\% (split-static; located 77.2\%), the price of a byte-mechanically verifiable no-qrel-leak guarantee. The decomposition's per-query dynamic arm shows a per-query dynamic fence would recover +2.28 points (composite contrast); building a byte-verifiable per-query fence is engineering future work.
\item \textbf{Mechanism divergence is an observation from two protocols}, with corpus density as a hypothesis, not a law.
\item \textbf{Dense comparison scope}: CLERC has a controlled \textbf{zero-shot} dense comparator (bge-large-en-v1.5, 512-token truncation affecting 29.5\% of passages; plus a frozen BM25+dense hybrid); ECtHR-PCR does not (published dense rows remain reference only). We make no paired trained-dense comparison on either protocol; the LeCoPCR comparison is published-rows only.
\item \textbf{Attempted trained-checkpoint reproduction (negative, reported for transparency)}: we attempted to reproduce the published CLERC ft-LegalBERT-DPR result (68.5 R@1000) for the officially released checkpoint under a good-faith Tevatron-v1 reconstruction (final-layer [CLS], unnormalized dot product, tied encoder, no prefixes, 512 tokens, official passage collection, MaxP). We obtained 61.59, outside our pre-registered tolerance; the public checkpoint is released without a model card or encode recipe, so the responsible inference detail cannot be identified. We do not read this as evidence against the published number and make no head-to-head claim.
\item \textbf{Operational footprint}: the CLERC anchor channel adds 16.57M anchor-window edges over 179 per-year Lucene indexes on a 1.84M-document corpus; it has no learned parameters and is BM25-class infrastructure (no GPU at index or query time), but we do not report latency/throughput measurements; a full cost profile accompanies the artifact release.
\item \textbf{Citation-derived relevance}: both benchmarks' labels are judicial citations; degree controls address the popularity confound within this regime, but the evaluation measures the ability to recover citation behavior, not independently adjudicated relevance.
\item \textbf{Window locating on ECtHR-PCR} uses digit-bound application-number matches and no external identity linker; in the frozen 100-window precision grading both failures are 1--2-digit serials colliding with dates or article numbers. Leakage numbers are method-internal (our fence-less variants), not measurements of prior systems.
\end{itemize}

\begin{table*}[t]
\centering
\footnotesize
\setlength{\tabcolsep}{3pt}
\begin{tabular}{cp{0.19\textwidth}p{0.13\textwidth}p{0.30\textwidth}p{0.22\textwidth}}
\toprule
\# & Read & Known at freeze & Frozen before it & What changed \\
\midrule
1 & Original fenced-vs-sparse (label-assisted alignment) & dev results only & dev-tuned fenced config, metric & first exposure; alignment later found to read qrels (provenance defect) \\
2 & Audited rerun (qrel-free map, policy estimand) & \#1's +21.98 & byte-invariance-audited mapper + calibration; policy/estimand/decision rule; 2 post-failure amendments (feasible-recall denominator; self-reference-query exclusion) & primary $\rightarrow$ policy +16.11; +21.98 demoted to superseded \\
3 & Zero-shot dense comparator & \#1--2 & dense encoder + hybrid config & reference comparison only \\
4 & Provenance/family-flag audit & \#1--2 & audit spec & 9/593 flip collapse after family-flag removal \\
5 & Fairly-tuned degree control & \#1--4 incl.\ +16.11 & 51-config grid, selection rule, seed, sealed before read & dev-optimal degree weight = 0 $\rightarrow$ control = BM25 \\
\bottomrule
\end{tabular}
\caption{Every CLERC test-qrel read, in order (each frozen before its own read; the test set was exposed at read \#1, so all later reads are post-exposure by construction).}
\label{tab:reads}
\end{table*}

\section{Conclusion}

Incoming-citation context is a strong retrieval signal for case law, but only once it is fenced in time; across two jurisdictions and two protocols we establish both halves of that claim. The fenced gain is large: a policy +16.1 R@1000 on the eligible CLERC test universe under a byte-invariance-audited mapping (+22.3 coverage-conditioned on mapped queries; policy LB95 +14.54), clearing a fairly-tuned citation-degree control whose dev-optimal degree weight is zero; and +8.68 macro R@1000 over BM25+degree on ECtHR-PCR (larger at shallow cuts, R@100 +15.45), at zero training. The fence-less version of the same method over-credits itself in both, and we can now say by how much and of what kind: on ECtHR-PCR, +2.70 points of genuinely non-prior-date evidence inside a naive +4.97 relax effect (54\% of it), but only 14.9\% of the full +18.1 citation-context gain over BM25; on CLERC, a dose--response climbing from 0 to +4.95 points, monotone in aggregate point estimates, as the year fence is relaxed. What the signal \emph{is} appears to differ by regime (consistent with candidate-specific content on CLERC and with fenced citation-window score mass on ECtHR-PCR, though the two placebo designs are not matched, so this cross-corpus contrast is observational), but what evaluation must do does not: (i) admit only citers dated before the query (and, where labels are citations, before the evaluation split); (ii) control for citation degree, not just a text baseline; (iii) report the fenced gain separately from the admission premium, named at the granularity the corpus dates support; (iv) disclose alignment/coverage. This paper's decompositions make each of these checkable.

\section*{Ethics Statement}

Our method performs first-stage \emph{retrieval}: it surfaces candidate precedents for a human reader and neither predicts case outcomes nor automates any judicial decision. Retrieved cases are suggestions to be verified by a qualified professional, not a substitute for legal judgment.

Both benchmarks are built from publicly released court documents---U.S.\ federal opinions (CLERC) and European Court of Human Rights judgments (ECtHR-PCR)---in their published form. We collect no new data, use no crowdsourced annotation, and introduce no private or personally identifying information beyond what these public corpora already contain; the single manual step, a 100-window precision grading, was carried out by the authors.

Relevance in both benchmarks is defined by judicial citation, so a system fit to recover citation behavior can inherit the biases of citation practice, such as favoring already well-cited or higher-court cases. This popularity confound is exactly what our degree and count controls target and part of our motivation: an unfenced or degree-uncontrolled citation-context signal can overstate a system's usefulness and encode such biases, while the temporal fence removes information no deployed system could have observed. We caution that our findings are specific to two jurisdictions and to citation-defined relevance and should not be assumed to transfer elsewhere without further validation.

\bibliography{custom}

@inproceedings{patil2024pat,
  title     = {Citation Anchor Text for Improving Precedent Retrieval: An Experimental Study on Indian Legal Documents},
  author    = {Patil, Gaurang and Sisodiya, Bhoomeendra Singh and Reddy, P. Krishna and Santhy, K. V. K.},
  booktitle = {Legal Knowledge and Information Systems: JURIX 2024},
  series    = {Frontiers in Artificial Intelligence and Applications},
  publisher = {IOS Press},
  year      = {2024},
  doi       = {10.3233/FAIA241241}
}

@inproceedings{craswell2001anchor,
  title     = {Effective Site Finding using Link Anchor Information},
  author    = {Craswell, Nick and Hawking, David and Robertson, Stephen},
  booktitle = {Proceedings of the 24th Annual International ACM SIGIR Conference on Research and Development in Information Retrieval},
  year      = {2001},
  publisher = {ACM}
}

@inproceedings{ritchie2008citation,
  title     = {Comparing Citation Contexts for Information Retrieval},
  author    = {Ritchie, Anna and Robertson, Stephen and Teufel, Simone},
  booktitle = {Proceedings of the 17th ACM Conference on Information and Knowledge Management (CIKM)},
  year      = {2008},
  publisher = {ACM}
}

@inproceedings{hou2025clerc,
  title     = {{CLERC}: A Dataset for {U.S.} Legal Case Retrieval and Retrieval-Augmented Analysis Generation},
  author    = {Hou, Abe Bohan and Weller, Orion and Qin, Guanghui and Yang, Eugene and Lawrie, Dawn and Holzenberger, Nils and Blair-Stanek, Andrew and Van Durme, Benjamin},
  booktitle = {Findings of the Association for Computational Linguistics: NAACL 2025},
  year      = {2025},
  publisher = {Association for Computational Linguistics}
}

@inproceedings{joshi2023ucreat,
  title     = {{U-CREAT}: Unsupervised Case Retrieval using Events extr{A}c{T}ion},
  author    = {Joshi, Abhinav and Sharma, Akshat and Tanikella, Sai Kiran and Modi, Ashutosh},
  booktitle = {Proceedings of the 61st Annual Meeting of the Association for Computational Linguistics (Volume 1: Long Papers)},
  year      = {2023},
  publisher = {Association for Computational Linguistics}
}

@inproceedings{santosh2024ecthrpcr,
  title     = {{ECtHR-PCR}: A Dataset for Precedent Understanding and Prior Case Retrieval in the {European Court of Human Rights}},
  author    = {Santosh, T.Y.S.S and Haddad, Rashid Gustav and Grabmair, Matthias},
  booktitle = {Proceedings of the 2024 Joint International Conference on Computational Linguistics, Language Resources and Evaluation (LREC-COLING 2024)},
  year      = {2024}
}

@inproceedings{santosh2025lecopcr,
  title     = {{LeCoPCR}: Legal Concept-guided Prior Case Retrieval for {European Court of Human Rights} Cases},
  author    = {Santosh, T.Y.S.S and Olgu{\'i}n Nolasco, Isaac Misael and Grabmair, Matthias},
  booktitle = {Findings of the Association for Computational Linguistics: NAACL 2025},
  pages     = {1654--1661},
  year      = {2025},
  publisher = {Association for Computational Linguistics}
}

@inproceedings{mumford2025citnet,
  title     = {Context-Aware Citation Networks: A Human--{AI} Dataset, Analysis, and Tool},
  author    = {Mumford, Jack and Florimonte, Francesco and Atkinson, Katie and Dzehtsiarou, Kanstantsin},
  booktitle = {Legal Knowledge and Information Systems: JURIX 2025},
  series    = {Frontiers in Artificial Intelligence and Applications},
  publisher = {IOS Press},
  year      = {2025},
  doi       = {10.3233/FAIA251584}
}

@inproceedings{mumford2025goldilocks,
  title     = {Finding the {Goldilocks} Zone: Retrieving Citation Context},
  author    = {Mumford, Jack and Bareham, David and Atkinson, Katie and Marshall, Jeremy},
  booktitle = {Proceedings of the Twentieth International Conference on Artificial Intelligence and Law (ICAIL)},
  year      = {2025},
  publisher = {ACM},
  doi       = {10.1145/3769126.3769261}
}

@inproceedings{premasiri2025llmembed,
  title     = {{LLM}-based Embedders for Prior Case Retrieval},
  author    = {Premasiri, Damith and Ranasinghe, Tharindu and Mitkov, Ruslan},
  booktitle = {Proceedings of Recent Advances in Natural Language Processing (RANLP)},
  pages     = {980--988},
  year      = {2025}
}

@misc{goyal2026profile,
  title         = {Public Profile Matters: A Scalable Integrated Approach to Recommend Citations in the Wild},
  author        = {Goyal, Karan and Kukreja, Dikshant and Goyal, Vikram and Mohania, Mukesh},
  year          = {2026},
  eprint        = {2603.17361},
  archivePrefix = {arXiv}
}

@inproceedings{ellahib2026leakage,
  title         = {Temporal Leakage in Search-Engine Date-Filtered Web Retrieval: A Retrospective Forecasting Case Study},
  author        = {El Lahib, Ali and Xia, Ying-Jieh and Li, Zehan and Wang, Yuxuan and Pi, Xinyu},
  year          = {2026},
  booktitle     = {Proceedings of the 64th Annual Meeting of the Association for Computational Linguistics (Volume 2: Short Papers)},
  pages         = {787--795},
  eprint        = {2602.00758},
  archivePrefix = {arXiv}
}

@article{ji2023leakage,
  title   = {A Critical Study on Data Leakage in Recommender System Offline Evaluation},
  author  = {Ji, Yitong and Sun, Aixin and Zhang, Jie and Li, Chenliang},
  journal = {ACM Transactions on Information Systems},
  volume  = {41},
  number  = {3},
  year    = {2023}
}

@misc{choi2026daldall,
  title         = {{DALDALL}: Data Augmentation for Lexical and Semantic Diverse in Legal Domain by leveraging {LLM}-Persona},
  author        = {Choi, Janghyeok and Lee, Jaewon and Cho, Sungzoon},
  year          = {2026},
  eprint        = {2603.22765},
  archivePrefix = {arXiv}
}

@misc{ovcharov2026statute,
  title         = {Temporal Decay of Co-Citation Predictability: A 20-Year Statute Retrieval Benchmark from 396{M} {Ukrainian} Court Citations},
  author        = {Ovcharov, Volodymyr},
  year          = {2026},
  eprint        = {2605.17639},
  archivePrefix = {arXiv}
}

@inproceedings{yang2019neuralhype,
  title={Critically Examining the ``Neural Hype'': Weak Baselines and the Additivity of Effectiveness Gains from Neural Ranking Models},
  author={Yang, Wei and Lu, Kuang and Yang, Peilin and Lin, Jimmy},
  booktitle={Proceedings of the 42nd International ACM SIGIR Conference on Research and Development in Information Retrieval (SIGIR)},
  year={2019}
}

@inproceedings{thakur2021beir,
  title={{BEIR}: A Heterogeneous Benchmark for Zero-shot Evaluation of Information Retrieval Models},
  author={Thakur, Nandan and Reimers, Nils and R{\"u}ckl{\'e}, Andreas and Srivastava, Abhishek and Gurevych, Iryna},
  booktitle={Proceedings of the Neural Information Processing Systems Track on Datasets and Benchmarks (NeurIPS Datasets and Benchmarks)},
  year={2021}
}

\appendix

\section{Qualitative Example: recovering a leading precedent via incoming-citation context}
\label{app:casestudy}

A concrete recovery shows why incoming-citation context helps. Test query 37245/13 (Poland, 2020) concerns the length and justification of the applicant's detention on remand while charged with membership of an organised armed group; among its gold precedents is \emph{Ilijkov v.\ Bulgaria} (no.\ 33977/96, 2001), the leading authority that the severity of the anticipated sentence cannot on its own justify prolonged pre-trial detention. \emph{Ilijkov}'s own facts, however, concern customs fraud and forged export declarations---lexically remote from the query's armed-group and detention-length facts---so a body-text retriever has little of the query's vocabulary to match against \emph{Ilijkov}'s opinion. The matching signal instead comes from how a \emph{later} case describes \emph{Ilijkov} when citing it: a 2006 judgment (no.\ 23042/02) cites it with the anchor window ``\ldots{} the authorities relied on the reasonable suspicion that the applicant had committed the offences with which he had been charged and on the severity of the sentence\ldots'', language that shares the query's terms (\emph{charged}, \emph{committed}, \emph{severity of the sentence}) far more than \emph{Ilijkov}'s own text does. Because this citing case predates the 2020 query, the temporal fence admits its window, and the anchor channel surfaces \emph{Ilijkov} where the body channel alone ranks it low. The window was graded a true citation context in our precision audit (\S\ref{sec:echrtest}); this recovery is one of the 3{,}484 golds the fenced channel recovers over BM25+degree.

\section{Reproducibility}
\label{app:repro}

Code and the derived retrieval runs needed to reproduce every headline number are provided as supplementary material and will be released publicly. The CLERC first stage is disk-based BM25 (Anserini/Lucene) tuned on dev over a $(k_1,b)$ grid; the reported official BM25 reproduces the published 0.483 exactly (0.4826). The fairly-tuned citation-degree control is selected on dev over a frozen 51-configuration grid spanning decayed and undecayed distinct-citer degree signals, a body/degree fusion weight $\beta$ (including body-only and degree-only endpoints), and the RRF constant $k$, with the winner evaluated once under the frozen policy estimand. The ECtHR-PCR anchor channel and its degree, count, and candidate-identity controls use the same RRF fusion; every placebo swap, cluster-bootstrap resampling ($B{=}10{,}000$), and the single test-qrel read use seeds fixed in advance and recorded in the release. The source-mapping procedure, the recency-decay scoring used for the dose--response sensitivity, the family-closure filter, and the fence-violation and byte-invariance audit routines are included so that the mapper, the decay formula, and the audit inputs can be inspected and re-run.

\section{COLIEE as a Negative Applicability Boundary (Probe)}
\label{app:coliee}

Incoming-citation-context methods need (a) in-corpus citing text, (b) mention$\rightarrow$document linkability, (c) temporal metadata. A measurement probe on COLIEE 2026 Task 1 shows that FRAGMENT\_SUPPRESSED appears in 1,813/1,848 test files and 7,308/7,708 train files, with 35,434 test and 214,824 train suppressed fragments. Surviving full-form citations are only 113 test / 375 train, outnumbered by suppressed fragments 314:1 / 573:1. Only 33/1,848 test files have any head-300 self-id-like pattern under the probe heuristic. Dates are extractable but noisy, while citation-window text and mention-to-document linkability are broken. Anchor-style enrichment is therefore possible only by importing external identities/graphs --- a resource-regime change, not a same-protocol replication. We document this as a protocol-selection boundary for the method family.

\section{Estimand Glossary}
\label{app:glossary}

\begin{itemize}
\item \textbf{Fenced gain}: fenced system $-$ designated primary comparator, same protocol (CLERC: tuned BM25; ECtHR-PCR: BM25+degree).
\item \textbf{Non-prior-date premium} (ECtHR-PCR; day granularity): unrestricted $-$ per-query dynamic admission ($c{-}b$), same scoring --- evidence dated on or after the query date.
\item \textbf{Non-prior-year admission premium} (CLERC; year granularity): unfenced $-$ prior-year fence, same scoring; includes same-calendar-year citers whose temporal direction is not identifiable from year-level dates.
\item \textbf{Split-static admission cost} (ECtHR-PCR): per-query dynamic $-$ split-static admission on the same artifact set ($b{-}a'$) --- legitimate pre-query evidence a split-level fence excludes.
\item \textbf{Index/retrieval-set effect} (ECtHR-PCR): split-static admission on full artifacts $-$ deployed split-static ($a'{-}a$); depth-sensitive, no sign claim.
\item \textbf{Phantom fraction} (ECtHR-PCR): $(c{-}b)/(c{-}\text{body})$, jointly cluster-resampled; the share of the naive unfenced-over-BM25 gain that is non-prior-date evidence.
\item \textbf{Dose--response premium($k$)} (CLERC): R@1000 under admission citer-year $< Y{+}k$ minus the $k{=}0$ fence, time-blind scoring frozen across arms.
\end{itemize}

\section{Dose--Response Per-Arm Values}
\label{app:dose}

\begin{table}[h]
\centering
\small
\setlength{\tabcolsep}{3.5pt}
\begin{tabular}{lrrcc}
\toprule
Arm & R@1000 & Prem. & Pointwise CI & Sim.\ band \\
\midrule
$k{=}0$ & 88.35 & 0 & --- & --- \\
$k{=}1$ & 89.75 & +1.40 & [0.90, 1.95] & [0.75, 2.05] \\
$k{=}2$ & 90.35 & +2.00 & [1.40, 2.60] & [1.23, 2.77] \\
$k{=}5$ & 91.70 & +3.35 & [2.60, 4.15] & [2.39, 4.31] \\
$k{=}10$ & 92.65 & +4.30 & [3.45, 5.20] & [3.21, 5.39] \\
$k{=}\infty$ & 93.30 & +4.95 & [4.00, 5.95] & [3.72, 6.18] \\
\bottomrule
\end{tabular}
\caption{CLERC dev fence-offset dose--response per-arm values (Figure~\ref{fig:dose}; time-blind scoring; premium in R@1000 points vs the $k{=}0$ fence; $B{=}10{,}000$ cluster bootstrap; sup-$t$ simultaneous bands). Admission: $k{=}0$ strict fence (citer-year $<$ Y); $k{=}1$ adds the same calendar year; $k{=}2$ adds year Y+1; $k{=}5$ adds Y+2..Y+4; $k{=}10$ adds Y+5..Y+9; $k{=}\infty$ removes the fence.}
\end{table}

\end{document}